\documentclass[12pt]{iopart}

\usepackage{graphicx}

\begin{document}

\title[Solar system tests of brane world models]
{Solar system tests of brane world models}

\author{Christian G. B\"ohmer\footnote[1]{c.boehmer@ucl.ac.uk}}
\address{Department of Mathematics, University College London,
Gower Street, London, WC1E 6BT, UK}

\author{Tiberiu Harko\footnote[2]{harko@hkucc.hku.hk}}
\address{Department of Physics and Center for Theoretical and
             Computational Physics, The University of Hong Kong,
             Pok Fu Lam Road, Hong Kong}

\author{Francisco S. N. Lobo\footnote[3]{francisco.lobo@port.ac.uk}}
\address{Institute of Cosmology \& Gravitation,
             University of Portsmouth, Portsmouth PO1 2EG, UK}
\address{Centro de Astronomia e Astrof\'{\i}sica da Universidade
             de Lisboa, Campo Grande, Ed. C8 1749-016 Lisboa, Portugal}


\begin{abstract}

The classical tests of general relativity (perihelion precession,
deflection of light, and the radar echo delay) are considered for
the Dadhich, Maartens, Papadopoulos and Rezania (DMPR) solution of
the spherically symmetric static vacuum field equations in
brane world models. For this solution the metric in the vacuum
exterior to a brane world star is similar to the
Reissner-Nordstrom form of classical general relativity, with the
role of the charge played by the tidal effects arising from projections
of the fifth dimension. The existing observational solar system data on
the perihelion shift of Mercury, on the light bending around the
Sun (obtained using long-baseline radio interferometry), and
ranging to Mars using the Viking lander, constrain the numerical
values of the bulk tidal parameter and of the brane tension.

\end{abstract}


\maketitle

\section{Introduction}

The idea that our four-dimensional Universe might be a three-brane
\cite{RS99a}, embedded in a higher dimensional space-time and
inspired by superstring theory, has recently attracted much
attention. In this context, the ten-dimensional $E_{8}\times
E_{8}$ heterotic string theory, which contains the standard model
of elementary particles, could be a promising candidate for the
description of the real Universe. This theory is connected with an
eleven-dimensional theory, the $M$-theory, compactified on the
orbifold $R^{10}\times S^{1}/Z_{2}$ \cite{HW96}. According to the
brane-world scenario, the physical fields in our four-dimensional
space-time, which are assumed to arise as fluctuations of branes
in string theories, are confined to the three brane. Only gravity
can freely propagate in the bulk space-time, with the
gravitational self-couplings not significantly modified. The model
originated from the study of a single $3$-brane embedded in five
dimensions, with the $5D$ metric given by $ds^{2}=e^{-f(y)}\eta
_{\mu \nu }dx^{\mu }dx^{\nu }+dy^{2}$, which due to the appearance
of the warp factor, could produce a large hierarchy between the
scale of particle physics and gravity. Even if the fifth dimension
is uncompactified, standard $4D$ gravity is reproduced on the
brane. Hence this model allows the presence of large, or even
infinite non-compact extra dimensions. Our brane is identified as
a domain wall in a $5$-dimensional anti-de Sitter space-time. For
a review of dynamics and geometry of brane Universes see
\cite{Ma01}.

Due to the correction terms coming from the extra dimensions,
significant deviations from the Einstein theory occur in brane
world models at very high energies \cite{SMS00,SSM00}. In
particular, gravity is largely modified at the electro-weak scale,
1TeV. The cosmological and astrophysical implications of the brane
world theories have been extensively investigated in the
literature \cite{all}. Gravitational collapse can also produce
high energies, with the five dimensional effects playing an
important role in the formation of black holes \cite{all1}.

Note that for standard general relativistic spherical compact
objects the exterior space-time is usually described by the
Schwarzschild metric. However, in five dimensional brane world
models, the high energy corrections to the energy density,
together with Weyl stresses from bulk gravitons, imply that on
the brane the exterior metric of a static star is no longer the
Schwarzschild metric \cite{Da00}. The presence of the Weyl
stresses also mean that the matching conditions do not have a
unique solution on the brane; the knowledge of the
five-dimensional Weyl tensor is needed as a minimum condition for
uniqueness.

Static and spherically symmetric exterior vacuum solutions of the
brane world models were initially proposed by Dadhich et al.
\cite{Da00} and Germani and Maartens \cite{GeMa01}. The former
solutions has the mathematical form of the Reissner-Nordstrom
solution, in which a tidal Weyl parameter plays the role of the
electric charge of the general relativistic solution  \cite{Da00}.
The solution was obtained by imposing the null energy condition on
the 3-brane for a bulk having non-zero Weyl curvature, and a
specific case was obtained by matching to an interior solution,
corresponding to a constant density brane world star. An exact
interior uniform-density stellar solution on the brane was also
found in \cite{GeMa01}. In the latter model, it was found that the
general relativistic upper bound for the mass-radius ratio,
$M/R<4/9$, was reduced by 5-dimensional high-energy effects. It
was also found that the existence of brane world neutron stars
leads to a constraint on the brane tension, which is stronger than
the big-bang nucleosynthesis constraint, but weaker than the
Newton-law experimental constraints \cite{GeMa01}. We refer the
reader to \cite{branesolutions1,branesolutions2} and references
therein for further static and spherically symmetric brane world
solutions.

There are several possibilities of observationally testing the
brane world models at an astrophysical/cosmological scale, such as
using the time delay of gamma ray bursts \cite{HaE} or by using
the luminosity distance--redshift relation  for supernovae at
higher redshifts \cite{GeE}. The classical tests of general
relativity, namely, light deflection, time delay and perihelion
shift, have been analyzed, for gravitational theories with large
non-compactified extra-dimensions, in the framework of the
five-dimensional extension of the Kaluza-Klein theory, using an
analogue of the four-dimensional Schwarzschild metric in
\cite{Li00}. Solar system data also imposes some strong
constraints on Kaluza-Klein type theories. The existence of
extra-dimensions and of the brane-world models can also be tested
via the gravitational radiation coming from primordial black
holes, with masses of the order of the lunar mass, $M \sim
10^{-7}M_{\odot }$, which might have been produced when the
temperature of the universe was around 1TeV. If a significant
fraction of the dark halo of our galaxy consists of these lunar
mass black holes, a huge number of black hole binaries could
exist. The detection of the gravitational waves from these
binaries could confirm the existence of extra-dimensions
\cite{In03}.

It is the purpose of the present paper to consider the classical
tests (perihelion precession, light bending and radar echo delay)
of general relativity for static gravitational fields in the
framework of brane world models. To do this we shall adopt for the
geometry of the brane outside a compact, stellar type object, the
spherically symmetric, Reissner-Nordstrom type, static brane
metric obtained by Dadhich, Maartens, Papadopoulos and Rezania
(DMPR for short) \cite{Da00}. For this metric, we first consider
the motion of a particle (planet), and the contributions of the
five-dimensional effects to the perihelion precession are
calculated.

By considering the motion of a photon in the static brane
gravitational field we obtain the corrections, due to the
projected bulk Weyl tensor, to the bending of light by massive
astrophysical objects and to the radar echo delay, respectively.
Existing data on light-bending around the Sun, using long-baseline
radio interferometry, ranging to Mars using the Viking lander, and
the perihelion precession of Mercury, can all give significant and
detectable solar system constraints associated with the
extra-dimensional part of the metric. More exactly, the study of
the classical general relativistic tests, by taking into account
the corrections coming from the extra dimensions, constrain the
tidal bulk parameter and, via the junction conditions, the brane
tension. The advance of the perihelion for the (charged) standard
general relativistic Reissner-Nordstrom metric has been considered
in \cite{Ch01}, where the formula for the shift in the perihelion
of a charged particle has been derived. The gravitational lensing
by a Reissner-Nordstrom black hole in the weak field limit has
also been analysed in \cite{Sereno:2003nd}.

The present paper is organized as follows. The static and
spherically symmetric vacuum solution on the brane is presented in
Section \ref{Sect2}. In Section \ref{Sect3} we consider the
classical solar system tests, namely, the perihelion shift, the
light deflection and the radar echo delay, for the brane world
model stars. We conclude our results in Section \ref{Sect4}.

\section{The DMPR solution of the vacuum field equations on the brane}\label{Sect2}

The $5$-dimensional Einstein field equation in the bulk is given
by $G_{IJ}=k_{5}^{2}\,T_{IJ}$, where the 5-dimensional
energy-momentum, $T_{IJ}$, is provided by
\begin{equation}
T_{IJ}=-\Lambda _{5}g_{IJ}+\delta (Y)\left[ -\lambda
g_{IJ}+T_{IJ}^{\rm matter}\right] ,
\end{equation}
and $\Lambda _{5}$ is the negative vacuum energy in the bulk. In
the analysis below, we consider that capital Latin indices run in
the range $0,...,4$, while Greek indices take the values
$0,...,3$.

The effective four-dimensional field equations on the brane (the
Gauss equation), take the form \cite{SMS00,SSM00}:
\begin{equation}
G_{\mu \nu }=-\Lambda g_{\mu \nu }+k_{4}^{2}T_{\mu \nu
}+k_{5}^{4}S_{\mu \nu }-E_{\mu \nu },  \label{Ein}
\end{equation}
where the four-dimensional cosmological constant, $\Lambda $, and
the coupling constant, $k_{4}$, are given by $\Lambda
=k_{5}^{2}\left( \Lambda _{5}+k_{5}^{2}\lambda ^{2}/6\right) /2$
and $k_{4}^{2}=k_{5}^{4}\lambda /6$. In the limit $\lambda
^{-1}\rightarrow 0$ we recover standard general relativity,
respectively, with $\lambda$ the vacuum energy on the brane.

$S_{\mu \nu }$ is the local quadratic energy-momentum correction,
which arises from the extrinsic curvature term in the projected
Einstein tensor, and is given by
\begin{equation}
S_{\mu \nu }=\frac{1}{12}TT_{\mu \nu }-\frac{1}{4}T_{\mu
}{}^{\alpha }T_{\nu \alpha }+\frac{1}{24}g_{\mu \nu }\left(
3T^{\alpha \beta }T_{\alpha \beta }-T^{2}\right) .
\end{equation}
The term $E_{\mu \nu }$ is the projection of the 5-dimensional
Weyl tensor $C_{IAJB}$, $E_{IJ}=C_{IAJB}\,n^{A}n^{B}$. The only
known property of this nonlocal term is that it is traceless, i.e,
$E^{\mu}{}_{\mu}=0$.

The Einstein equation in the bulk and the Codazzi equation, also
imply the conservation of the energy-momentum tensor of matter on
the brane, $D_{\nu }T_{\mu }{}^{\nu }=0$, where $D_{\nu }$ denotes
the brane covariant derivative. Moreover, the contracted Bianchi
identities on the brane imply that the projected Weyl tensor
should obey the constraint $D_{\nu }E_{\mu }{}^{\nu
}=k_{5}^{4}D_{\nu }S_{\mu }{}^{\nu }$.

Note that the symmetry properties of $E_{\mu \nu }$ imply that in
general we can decompose it irreducibly with respect to a chosen
$4$-velocity field $u^{\mu }$ as
\begin{equation}
E_{\mu \nu }=-\tilde{k}^{4}\left[ U\left( u_{\mu }u_{\nu
}+\frac{1}{3}h_{\mu \nu }\right) +2Q_{(\mu }u_{\nu )}+P_{\mu \nu
}\right] ,  \label{WT}
\end{equation}
where $\tilde{k}=k_{5}/k_{4}$, $h_{\mu \nu }=g_{\mu \nu }+u_{\mu
}u_{\nu }$
projects orthogonal to $u^{\mu }$, the ``dark radiation'' term $U=-\tilde{k}%
^{-4}E_{\mu \nu }u^{\mu }u^{\nu }$ is a scalar, $Q_{\mu
}=\tilde{k}^{-4}h_{\mu
}^{\alpha }E_{\alpha \beta }u^{\beta }$ a spatial vector and $P_{\mu \nu }=-\tilde{k}%
^{-4}\left[ h_{(\mu }^{\alpha }h_{\nu )}^{\beta }-\frac{1}{3}%
h_{\mu \nu }h^{\alpha \beta }\right] E_{\alpha \beta }$ a spatial,
symmetric and trace-free tensor \cite{Ma01}.

For the specific case of vacuum, $T_{\mu \nu }=0$, and
consequently $S_{\mu \nu }=0$, and assuming that $\Lambda=0$, the
field equations describing a static brane take the form
\begin{equation}
R_{\mu \nu }=-E_{\mu \nu },
\end{equation}
with $R_{\mu }^{\mu }=0=E_{\mu }^{\mu }$. For this case $E_{\mu
\nu }$ satisfies the constraint $D_{\nu }E_{\mu }{}^{\nu }=0$. In
a static vacuum $Q_{\mu }=0$ and the constraint for $E_{\mu \nu }$
takes the form
\begin{equation}
\frac{1}{3}D_{\mu }U+\frac{4}{3}UA_{\mu }+D^{\nu }P_{\mu \nu }+A^{\nu
}P_{\mu \nu }=0,
\end{equation}
where $D_{\mu }$ is the projection (orthogonal to $u^{\mu }$) of
the covariant derivative and $A_{\mu }=u^{\nu }D_{\nu }u_{\mu }$
is the
4-acceleration.

In the static spherically symmetric case we may choose $%
A_{\mu }=A(r)r_{\mu }$ and $P_{\mu \nu }=P(r)\left( r_{\mu }r_{\nu }-\frac{1%
}{3}h_{\mu \nu }\right) $, where $A(r)$ and $P(r)$ are some scalar
functions of the radial distance $r$, and $r_{\mu }$ is a unit
radial vector. The choice $U=\tilde{k}^{4}Q/r^{4}=-P/2$, where $Q$
is a constant, leads to a Reissner-Nordstrom type solution of the
static, spherically symmetric field equations on the brane
\cite{Da00}:
\begin{equation}
\fl ds^{2}=-\left( 1-\frac{2M}{M_{p}^{2}}\frac{1}{r}+\frac{q}{\tilde{M}_{p}^{2}}%
\frac{1}{r^{2}}\right) dt^{2}+\frac{dr^{2}}{1-\frac{2M}{M_{p}^{2}}\frac{1}{r}%
+\frac{q}{\tilde{M}_{p}^{2}}\frac{1}{r^{2}}}+r^{2}\left( d\theta ^{2}+\sin
^{2}\theta d\varphi ^{2}\right) ,  \label{metr}
\end{equation}
where the Planck scales in the brane and in the bulk, $M_{p}$ and $\tilde{M}%
_{p}$, respectively, are related by $M_{p}=\sqrt{3/4\pi }\tilde{%
M}_{p}^{3}/\sqrt{\lambda }$, and $q=Q\tilde{M}_{p}^{2}$ is the
dimensionless tidal parameter.

For this model the expression of the projected Weyl tensor,
transmitting the tidal charge stresses from the bulk to the brane
is \cite {Da00}
\begin{equation}
E_{\mu \nu }=-\frac{Q}{r^{4}}\left( u_{\mu }u_{\nu }-2r_{\mu }r_{\nu
}+h_{\mu \nu }\right) .
\end{equation}

Perturbative studies of the static weak-field regime show that the
leading order correction to the Newtonian potential on the brane
is given by $\phi =GM\left( 1+2l^{2}/3r^{2}\right) /r$, where $l$
is the curvature scale of the five-dimensional anti de Sitter
space time (AdS$_{5}$) \cite{RS99a}. However, this result assumes
that the bulk perturbations are bounded in conformally Minkowski
coordinates and that the bulk is nearly AdS$_{5}$ \cite{GeMa01}.
Different bulk geometries could induce different corrections to
Newton's law on the brane. In the following we denote by
$r_{g}=2M/M_{p}^{2}=2GM/c^{2}$ the gravitational radius of the
brane star.

\section{The classical tests of general relativity for the DMPR brane world solution}
\label{Sect3}

To determine the trajectory of a massive particle in the metric
(\ref{metr}) we use the Hamilton-Jacobi equation,
\begin{equation}
g^{ik}\frac{\partial S}{\partial x^{i}}\frac{\partial S}{\partial x^{k}}%
-m^{2}c^{2}=0,  \label{hj}
\end{equation}
where $m\neq 0$ is the mass of the particle \cite{LaLi}. As in
every central spherically symmetric field, the motion occurs in a
single plane passing through the origin, and without a loss of
generality one may choose the plane with $\theta =\pi/2$. With the
use of the metric coefficients from Eq. (\ref{metr}), we obtain
\begin{equation}
\fl \left( 1-\frac{r_{g}}{r}+\frac{Q}{r^{2}}\right) ^{-1}\left( \frac{\partial S%
}{c\partial t}\right) ^{2}-\left( 1-\frac{r_{g}}{r}+\frac{Q}{r^{2}}\right)
\left( \frac{\partial S}{\partial r}\right) ^{2}-\frac{1}{r^{2}}\left( \frac{%
\partial S}{\partial \varphi }\right) ^{2}-m^{2}c^{2}=0.  \label{hj1}
\end{equation}

According to the standard procedure for solving the
Hamilton-Jacobi equation \cite{La76}, we chose $S$ in the form
\begin{equation}
S=-Et+L\varphi +S_{r}\left( r\right) ,  \label{sol}
\end{equation}
where the energy $E$ and the angular momentum $L$ are constants of
motion. Substituting Eq. (\ref{sol}) into Eq. (\ref{hj1}), we find
$S_{r}(r)$ given by
\begin{equation}
S_{r}=\int \sqrt{\frac{E^{2}}{c^{2}}\left( 1-\frac{r_{g}}{r}+\frac{Q}{r^{2}}%
\right) ^{-2}-\left( m^{2}c^{2}+\frac{L^{2}}{r^{2}}\right) \left( 1-\frac{%
r_{g}}{r}+\frac{Q}{r^{2}}\right) ^{-1}}\;dr.  \label{sr}
\end{equation}

Considering a change in the integration variable from $r$ to
$r^{\prime }$ by means of the following transformation
\begin{equation}
r\left( r-r_{g}\right) +Q=r^{\prime 2},  \label{trans1}
\end{equation}
provides
\begin{equation}
\frac{r}{r^{\prime }}\approx 1+\frac{r_{g}}{2r^{\prime }}+\frac{r_{g}^{2}}{%
8r^{\prime 2}}-\frac{Q}{2r^{\prime 2}}.  \label{trans2}
\end{equation}

In terms of the new variable, and also by introducing the
non-relativistic energy $E_{0}$ ($E=E_{0}+mc^{2}$), we obtain
$S_{r}$ in the form
\begin{eqnarray}
S_{r}&=&\int \Bigg[\left( 2E_{0}m+\frac{E_{0}^{2}}{c^{2}}\right)
+\frac{1}{r} \left( 4E_{0}mr_{g}+2Gm^{2}M\right)
    \nonumber  \\
&&-\frac{L^{2}}{r^{2}}\left( 1-\frac{
3m^{2}c^{2}r_{g}^{2}-2m^{2}c^{2}Q}{2L^{2}}\right)
\Bigg]^{1/2}\;dr, \label{sr1}
\end{eqnarray}
where for brevity the prime on $r^{\prime }$ has been dropped.

\subsection{The precession of the perihelion}

The trajectory of the particle is defined by the equation $\varphi
+\left( \partial S_{r}/\partial L\right) ={\rm constant}$
\cite{La76}. Hence, a change of the angle $\varphi $ after one
revolution of the particle in the orbit is given by $\Delta
\varphi =-\left( \partial \Delta S_{r}/\partial L\right) $
\cite{LaLi}. Expanding $S_{r}$ in powers of the small correction
to the coefficient of $1/r^{2}$ we obtain
\begin{equation}
\Delta S_{r}=\Delta S_{r}^{(0)}-\frac{3m^{2}c^{2}r_{g}^{2}-2m^{2}c^{2}Q}{4L}%
\frac{\partial }{\partial L}\Delta S_{r}^{(0)},
\end{equation}
where $\Delta S_{r}^{(0)}$ corresponds to the motion in the closed
(Newtonian and unshifted) ellipse. Differentiating this relation
with respect to $L$, and taking into account that $\Delta \varphi
^{(0)}=-\left(
\partial \Delta S_{r}^{(0)}/\partial L\right) =2\pi $, we find
\begin{equation}
\Delta \varphi =2\pi \left(
1+\frac{3m^{2}c^{2}r_{g}^{2}-2m^{2}c^{2}Q}{4L^{2} }\right) .
\end{equation}

With the use of the Newtonian relation between the angular
momentum, the length of the semi-major axis $a$ and the
eccentricity $e$ of the ellipse, $ L^{2}/GMm^{2}=a\left(
1-e^{2}\right) $ \cite{La76}, we obtain the final form of the
precession $\delta \varphi =\Delta \varphi -2\pi $ of the
perihelion of a planet moving in the static gravitational field on
the DMPR brane:
\begin{equation}
\delta \varphi =\frac{6\pi GM}{c^{2}a\left( 1-e^{2}\right)
}-\frac{\pi c^{2}Q }{GMa\left( 1-e^{2}\right) }.  \label{del}
\end{equation}

The first term in Eq. (\ref{del}) is the well-known general
relativistic correction term for the perihelion precession, while
the second term gives the correction due to the non-local effects
arising from the Weyl curvature in the bulk.

The observed value of the perihelion precession of the planet Mercury is $%
\delta \varphi _{Obs}=43.11\pm 0.21$ arcsec per century
\cite{Sha76}. The general relativistic formula for the precession,
$\delta \varphi _{GR}=6\pi GM/c^{2}a\left( 1-e^{2}\right) $, with
$M=M_{\odot }=1.989\times 10^{33}$ g,
$c=2.998\times 10^{10}$ cm/s, $G=6.67\times 10^{-8}$ cm$^{3}$g$^{-1}$s$^{-2}$%
, $a=57.91\times 10^{11}$ cm and $e=0.205615$ \cite{Sha76}, gives
$\delta \varphi _{GR}=42.94$ arcsec per century. Therefore, the
difference $\Delta \varphi =\delta \varphi _{Obs}-\delta \varphi
_{GR}=0.17$ arcsec per century can be attributed to other effects.
By assuming that $\Delta \varphi $ is entirely due to the
modifications of the general relativistic Schwarzschild geometry
as a result of the five dimensional bulk effects, the
observational
results impose the following general constraint on the bulk tidal parameter $%
Q$:
\begin{equation}
\left| Q\right| \leq \frac{GM_{\odot }a\left( 1-e^{2}\right) }{\pi c^{2}}%
\Delta \varphi .  \label{Qobs}
\end{equation}

With the use of the observational data for Mercury, Eq.
(\ref{Qobs}) gives $ \left| Q\right| \leq 5.17\times 10^{8}$
cm$^{2}$, or in the natural system of units, with $c=\hbar =G=1$,
$\left| Q\right| \leq 1.32\times 10^{30}$ MeV$^{-2}$. On the other
hand, for a constant density star, Germani and Maartens
\cite{GeMa01} have derived the matching conditions at the vacuum
boundary of the brane star, implying $Q=-\left( 3GM/c^{2}\right)
R\left( \rho /\lambda \right) $, where $\rho $ is the density of
the brane star, $R$ its radius and $\lambda $ is the brane
tension. Therefore, by assuming that the Sun can be described (at
least approximatively) as a constant density brane star, we obtain
the following solar system observational constraint on the brane
tension $ \lambda $:
\begin{equation}
\lambda \geq \frac{3\pi R_{\odot }\rho _{\odot }}{a\left(
1-e^{2}\right) \Delta \varphi }. \label{l}
\end{equation}

The matching conditions for uniform density stars \cite{GeMa01}
also give
\begin{equation}
\lambda >\left[ GM/c^{2}/\left( R-2GM/c^{2}\right) \right] \rho
\,.
\end{equation}

For a typical neutron star with mass $M=1.4M_{\odot }$, density
$\rho =2\times 10^{14}$ g/cm$^{3}$ and $R=10\,$ km, we find
$\lambda
>7\times 10^{13}$ g/cm$^{3}( =2.905\times 10^{8}\;{\rm
MeV}^{4})$. By taking, for the case of the Sun, $R_{\odot
}=7\times 10^{10}\,$ cm and for $\rho$ the mean density of the
Sun, $\rho =\rho_{\odot }=1.41\,$ g/cm$^{3}$, Eq. (\ref{l}) gives,
with $\Delta \varphi =0.17$ arcsec per century, $\lambda \geq
8.4\times 10^{7}\, $g/cm$^{3}( =3.5\times 10^{2}\;{\rm MeV}^{4})$.

Unfortunately, the observational data on the perihelion precession
are strongly affected by the solar oblateness, whose value is
poorly known. Solar oblateness introduces a supplementary term of
the form $\xi J_{2}\delta \varphi _{GR}$ in the right hand side of
Eq. (\ref{del}), with $\xi =R_{\odot }^{2}/2M_{\odot }a\left(
1-e^{2}\right) $ and $J_{2}$ the solar quadrupole moment
\cite{Ca83}. The value of $J_{2}$ is a subject of debate, but a
recent estimate gives $J_{2}=\left( 3.64\pm 2.84\right) \times
10^{-6}$ \cite{RoRo97}. Taking into account the quadrupole
correction to $ \Delta \varphi $ gives an estimate of the brane
tension of the order $\lambda \geq 2\times 10^{10}$
g/cm$^{3}\left( =9\times 10^{4}{\rm MeV}^{4}\right)$.

\subsection{Light deflection on the brane}

The propagation of light in a centrally symmetric gravitational
field is described by the eikonal equation \cite{LaLi}
\begin{equation}
g^{ik}\frac{\partial \psi }{\partial x^{i}}\frac{\partial \psi }{\partial
x^{k}}=0.
\end{equation}

We assume again that the light ray is moving in the plane $\theta =\pi /2$.
By representing the eikonal $\psi $ in the form $\psi =-\omega
_{0}t+L\varphi +\psi _{r}\left( r\right) $, where $\omega _{0}$ is the
frequency of the light and $L$ a constant, it follows that the radial part
of the eikonal $\psi _{r}\left( r\right) $ is given by
\begin{equation}
\psi _{r}\left( r\right) =\frac{\omega _{0}}{c}\int \sqrt{\frac{r^{4}}{%
\left( r^{2}-r_{g}r+Q\right)
^{2}}-\frac{l^{2}}{r^{2}-r_{g}r+Q}}\;dr, \label{eik}
\end{equation}
where we denoted $l=cL/\omega _{0}$. By means of the transformations given
by Eqs. (\ref{trans1}) and (\ref{trans2}), Eq. (\ref{eik}) can be written as
\begin{equation}
\psi _{r}\left( r\right) =\frac{\omega _{0}}{c}\int
\sqrt{1+\frac{2r_{g}}{r}-\frac{l^{2}+2Q}{r^{2}}}\;dr.
\end{equation}

Expanding the integrand in powers of $r_{g}/r$ we obtain
\begin{equation}
\fl \psi _{r}=\psi _{r}^{(0)}+\frac{\omega _{0}r_{g}}{c}\int \frac{dr}{\sqrt{%
r^{2}-\left( l^{2}+2Q\right) }}=\psi _{r}^{(0)}+\frac{\omega _{0}r_{g}}{c}%
\cosh ^{-1}\frac{r}{\sqrt{l^{2}+2Q}},
\end{equation}
where $\psi _{r}^{(0)}$ corresponds to the classical straight ray,
with $r=l/\cos \varphi $. The total change in $\psi _{r}$ during
the propagation of the light from a very distant point $R$ to the
point $r=l$ nearest to the center and then back to $R$ is $\Delta
\psi
_{r}=\Delta \psi _{r}^{(0)}+(2\omega _{0}r_{g}/c)\cosh ^{-1}\left( R/\sqrt{%
l^{2}+2Q}\right) $.

The change in the polar angle is obtained by differentiating
$\Delta \psi _{r}$ with respect to $L$ \cite{LaLi}:
\begin{equation}
\Delta \varphi _{LD}=-\frac{\partial \Delta \psi _{r}}{\partial L}=-\frac{%
\partial \Delta \psi _{r}^{(0)}}{\partial L}+\frac{2r_{g}}{l}
\left( 1+\frac{2Q}{l^{2}}\right)^{-1} \left(
1-\frac{l^{2}+2Q}{R^{2}}\right) ^{-1/2}.
\end{equation}

Going to the limit $R\rightarrow \infty $ and taking into account that the
straight line corresponds to $\Delta \varphi =\pi $, we find that the angle $%
\delta \varphi _{LD}=\Delta \varphi _{LD}-\pi $ between the two asymptotes
of the light ray differs from $\pi $ by the angle
\begin{equation}
\delta \varphi _{LD}=\frac{2r_{g}}{l}-\frac{4Qr_{g}}{l^{3}}=\delta
\varphi _{LD}^{(GR)}\left( 1-\frac{2Q}{l^{2}}\right) ,
\end{equation}
where $\delta \varphi _{LD}^{(GR)}=4GM/c^{2}l$ is the standard
general relativistic light deflection term \cite{LaLi}.

We consider now the constraints on the brane world models arising
from the solar system observations of the light deflections. The
best available data come from long baseline radio interferometry
\cite{all2}, which gives $\delta \varphi _{LD}=\delta \varphi
_{LD}^{(GR)}\left( 1+\Delta _{LD}\right) $, with $\Delta _{LD}\leq
0.0017$. Therefore we have $\left| Q\right| \leq l^{2}\Delta
_{LD}/2$. For light just grazing the Sun's limb $l=R_{\odot }$,
and light deflection in the solar system imposes the restriction
$\left| Q\right| \leq 4\times 10^{18}\,{\rm cm}^{2}$ on the tidal
parameter describing the effect of the five dimensional bulk on
the brane.

Assuming again that the Sun can be modelled as a constant density
star, and taking into account the junction condition, we obtain
the following limit on the brane tension $\lambda$
\begin{equation}
\lambda \geq \frac{3GM}{c^{2}}\frac{\rho }{R\Delta _{LD}}\,.
\label{c2}
\end{equation}
As applied to the case of the Sun, Eq. (\ref{c2}) gives $\lambda
\geq 5.2\times 10^{-3}\,{\rm g/cm}^{3}(=2. 15\times 10^{-8}\,{\rm
MeV}^{4})$. However, if we admit that $\Delta _{LD}$ is a
universal quantity, giving the absolute deviation from standard
general relativity, we can equally apply Eq. (\ref{c2}) to high
density compact objects, such as neutron stars, by using the same
$\Delta _{LD}$ as obtained in the case of the solar system. With
$M=1.4M_{\odot}$, $\rho =2\times 10^{14}\,{\rm g/cm}^{3}$ and
$R=10^{6}\,{\rm cm}$, Eq. (\ref{c2}) gives $\lambda \geq 7.3\times
10^{16}\,{\rm g/cm}^{3}( =3\times 10^{11}\,{\rm MeV}^{4})$. The
light deflection for the DMPR black hole solution has also been
analysed in \cite{GD}, but using different methods.

\subsection{Radar echo delay}

A third solar system test of general relativity is the radar echo
delay \cite{Sh}. The idea of this test is to measure the time
required for radar signals to travel to an inner planet or
satellite in two circumstances: a) when the signal passes very
near the Sun and b) when the ray does not go near the Sun. The
null geodesic equation in the brane world metric (\ref{metr}) for
a radar signal travelling in the $\theta =\pi /2$ plane is
\begin{equation}
\left( \frac{dr}{cdt}\right) ^{2}=\left(
1-\frac{r_{g}}{r}+\frac{Q}{r^{2}} \right) ^{2}-r^{2}\left(
1-\frac{r_{g}}{r}+\frac{Q}{r^{2}}\right) \left( \frac{d\varphi
}{cdt}\right) ^{2}.  \label{ray}
\end{equation}

Note that the conservation of the angular momentum and energy
implies $r^{2}d\varphi /d\sigma ={\rm constant}$ and $\left(
1-r_{g}/r+Q/r^{2}\right) (cdt/d\sigma )={\rm constant}$, where
$\sigma $ is the parameter along the photon path, thus giving
$r^{2}\left( 1-r_{g}/r+Q/r^{2}\right)^{-1} (d\varphi /cdt)=b= {\rm
constant}$. Hence Eq. (\ref{ray}) becomes
\begin{equation}
\left( \frac{dr}{cdt}\right) ^{2}=\left( 1-\frac{r_{g}}{r}+\frac{Q}{r^{2}}%
\right) ^{2}-\frac{b^{2}}{r^{2}}\left( 1-\frac{r_{g}}{r}+\frac{Q}{r^{2}}%
\right) ^{3}.  \label{b}
\end{equation}

Let $PSE$ be the path of light from the planet, $P$, to the Earth,
$E$, with $S$ being the point of closest approach to the Sun. At
$P$, $S$ and $E$ the distances are $r=R_{P}$, $r=R_{S}$ and
$r=R_{E}$, respectively. The time taken by the light ray to travel
from $P$ to $E$ is \cite{Ra92}
\begin{equation}
t=\frac{1}{c}\sum_{i=P,E}\int_{R_{S}}^{R_{i}}\frac{dr}{\left( 1-\frac{r_{g}}{%
r}+\frac{Q}{r^{2}}\right) \sqrt{1-\frac{b^{2}}{r^{2}}\left( 1-\frac{r_{g}}{r}%
+\frac{Q}{r^{2}}\right) }}\,.
\end{equation}

Since $dr/dt=0$ at $r=R_{S}$ \cite{Ra92}, Eq. (\ref{b}) fixes the
value of the constant $b$ as $b^{2}=R_{S}^{2}/\left(
1-r_{g}/R_{S}+Q/R_{S}^{2}\right) $. With this value of $b$ and by
using the following approximation
\begin{equation}
\fl 1-b^{2}\left( 1-r_{g}/r+Q/r^{2}\right) /r^{2}\approx \left[
1-\left( R_{S}^{2}-Q\right) /r^{2}\right] \left[
1-R_{S}r_{g}/r\left( r+\sqrt{R_{S}^{2}-Q}\right) \right] \,,
\end{equation}
we obtain
\begin{equation}
t=\frac{1}{c}\sum_{i=P,E}\int_{R_{S}}^{R_{i}}\frac{1}{\sqrt{1-\frac{%
R_{S}^{2}-Q}{r^{2}}}}\left[ 1+\frac{R_{S}r_{g}}{2r\left( r+\sqrt{R_{S}^{2}-Q}%
\right) }+\frac{r_{g}}{r}-\frac{Q}{r^{2}}\right] dr.
\end{equation}
In the absence of the standard general relativistic gravitational
deflection of light, all terms in $r_{g}$ vanish and we obtain
\begin{equation}\label{t0}
t_{0}=\frac{1}{c}\sum_{i=P,E}\int_{R_{S}}^{R_{i}}\frac{1}{\sqrt{1-\frac{%
R_{S}^{2}-Q}{r^{2}}}}\left[ 1-\frac{Q}{r^{2}}\right] dr.
\end{equation}

In Eq. (\ref{t0}) we have assumed that the propagation of light in
the vacuum far away from matter sources is influenced by the bulk
effects only, characterized by the tidal coefficient $Q$. The
standard general relativistic effects are not present in this
approximation. Performing the integration gives
\begin{eqnarray}\label{tc}
\fl ct_{0} =\sqrt{R_{E}^{2}-R_{S}^{2}+Q}+\sqrt{R_{P}^{2}-R_{S}^{2}+Q}-2\sqrt{Q}%
    \nonumber\\
\fl +\frac{Q}{\sqrt{Q-R_{S}^{2}}}\ln \frac{R_{S}^{2}\left( \sqrt{Q-R_{S}^{2}}+%
\sqrt{R_{E}^{2}-R_{S}^{2}+Q}\right) \left( \sqrt{Q-R_{S}^{2}}+\sqrt{%
R_{P}^{2}-R_{S}^{2}+Q}\right) }{R_{E}R_{P}\left( \sqrt{Q}+\sqrt{Q-R_{S}^{2}}%
\right) ^{2}}.
\end{eqnarray}

Equation (\ref{tc}) describes the propagation of a light ray in a
brane world vacuum with a Reissner-Nordstrom type geometry induced
by the five-dimensional bulk effects. In order that $ct_{0}$ be a
real quantity it is necessary that the tidal coefficient $Q$ be
positive, $Q>0$. For $Q<0$, $t_0$ is an imaginary quantity. In the
limit $Q\rightarrow 0$ we obtain the classical (Newtonian) result
$ct_{0}^{(New)}=\sqrt{R_{E}^{2}-R_{S}^{2}}+\sqrt{R_{P}^{2}-R_{S}^{2}}$.

The time delay $2\left(t-t_0\right)$ for a round trip of a radar
signal travelling to a planet or satellite is
\begin{eqnarray}\label{td}
\fl 2\left( t-t_{0}\right)  \equiv \Delta
t_{RD}=2\frac{r_{g}}{c}\ln \frac{\left(
R_{E}+\sqrt{R_{E}^{2}-R_{S}^{2}+Q}\right) \left( R_{P}+\sqrt{%
R_{P}^{2}-R_{S}^{2}+Q}\right) }{\left( \sqrt{Q}+R_{S}\right) ^{2}}
 \nonumber\\
\fl +\frac{r_{g}}{c}\frac{R_{S}}{\sqrt{R_{S}^{2}-Q}}\left( \sqrt{\frac{R_{E}-%
\sqrt{R_{S}^{2}-Q}}{R_{E}+\sqrt{R_{S}^{2}-Q}}}+\sqrt{\frac{R_{P}-\sqrt{%
R_{S}^{2}-Q}}{R_{P}+\sqrt{R_{S}^{2}-Q}}}-2\sqrt{\frac{R_{S}-\sqrt{R_{S}^{2}-Q%
}}{R_{S}+\sqrt{R_{S}^{2}-Q}}}\right) .
\end{eqnarray}

In the limit $Q\rightarrow 0$ and by assuming that $R_{S}\ll
R_{E},R_{P}$ we recover the standard general relativistic
expression $\Delta t_{RD}^{(GR)}\approx
2r_{g}\left[ \ln \left( 4R_{E}R_{P}/R_{S}^{2}\right) +1\right] $ \cite{Sh}%
. The correction term $\Delta t_{BW}$ to the radar echo delay, due to the
effects of the five-dimensional bulk, can be written as
\begin{equation}
\Delta t_{RD}^{(BW)}\approx \frac{2r_{g}}{c}\left[ \ln \frac{\left( 1+\frac{Q%
}{4R_{E}^{2}}\right) \left( 1+\frac{Q}{4R_{P}^{2}}\right) }{\left( 1+\frac{%
\sqrt{Q}}{R_{S}}\right) ^{2}}+\frac{Q}{2R_{S}^{2}}-\sqrt{\frac{1-\sqrt{1-%
\frac{Q}{R_{S}^{2}}}}{1+\sqrt{1-\frac{Q}{R_{S}^{2}}}}}\right] .
\end{equation}

The total delay time for a radar signal travelling in the brane world is $%
\Delta t_{RD}=\Delta t_{RD}^{(GR)}+\Delta t_{RD}^{(BW)}$. However,
Eq. (\ref{td}) for the coordinate time delay is not very useful
directly, for it requires the knowledge of
$\left(R_{P}^2-R_{S}^2\right)^{1/2}$, etc, to a high degree of
accuracy. The differential systems of radial coordinates
themselves differ by large amounts. Besides, the electrons in the
solar corona affect the time delay by an amount which shows
considerable variation with time \cite{Ra92}.

The best experimental solar system constraints on time delay so
far have come from the Viking lander on Mars \cite{Re79}. In the
Viking mission two transponders landed on Mars and two others
continued to orbit round it. The latter two transmitted two
distinct bands of frequencies and thus the coronal effect could be
corrected for. For the time delay of the signals emitted on Earth
and which graze the Sun one obtains $\Delta t_{RD}=\Delta
t_{RD}^{(GR)}\left(1+\Delta _{RD}\right)$, with $\Delta _{RD}\leq
0.002$ \cite{Re79}. Therefore the contribution due to the five
dimensional bulk effects to the radar echo delay must satisfy the
constraint $\Delta t_{RD}^{(BW)}\leq \Delta _{RD}\Delta
t_{RD}^{(GR)}\leq 0.002\Delta t_{RD}^{(GR)}$.

For the case of the Earth-Mars-Sun system we have $R_E=1.525\times
10^{13}\,{\rm cm}$ (the distance Earth-Sun) and $R_P=2.491\times
10^{13}\,{\rm cm}$ (the distance Mars-Sun). With these values the
standard general relativistic radar echo delay has the value
$\Delta t_{RD}^{(GR)}\approx 2.68\times 10^{-4}\,{\rm s}$. Hence
the numerical value of the bulk tidal parameter $Q$ can be
constrained, by using radar echo delay data, via the relation
$\Delta t_{RD}^{(BW)}\leq 5.36\times 10^{-7}$, or equivalently,
\begin{eqnarray}
\ln \left[ \frac{\left( 1+1.07\times 10^{-27}Q\right) \left(
1+4.02\times 10^{-28}Q\right) }{\left( 1+1.43\times
10^{-9}\sqrt{Q}\right) ^{2}}\right]
     \nonumber    \\
+1.032\times 10^{-22}Q-7.18\times 10^{-12}\sqrt{Q} \left(
1+2.58\times 10^{-23}Q\right) \leq 0.027 \,.
\end{eqnarray}

By neglecting, in the first approximation, the logarithmic term
and taking into account that the term containing $\sqrt{Q}$
dominates, the radar echo delay results give the following general
restriction on the bulk tidal parameter:
\begin{equation}
\left| Q\right| \leq \frac{4c^{2}}{r_{g}^{2}}\left( \Delta
t_{RD}^{(GR)}\Delta _{RD}\right) ^{2}R_{S}^{2}.
\end{equation}
By using the numerical data we obtain $\left| Q\right| \leq 5.78\times
10^{19}$ cm$^{2}$, a value which is relatively consistent (taking also into
account the approximations we have used) with the similar value obtained
from the study of the deflection of light by using long baseline
interferometry data.

\section{Conclusion}\label{Sect4}

In the present paper, we considered the observational and
experimental possibilities for testing at the level of the solar
system the DMPR solution of the vacuum field equations in brane
world models. The classical tests of general relativity in the
solar system give strong constraints on the numerical values of
the brane tension and of the bulk tidal parameter, respectively.
Perihelion precession, light deflection and radar echo delay all
give definite constraints on the numerical value of $Q$. While the
two estimates obtained by using electromagnetic waves propagation
data (light deflection and radar delay) are relatively consistent
with each other, giving $\left| Q\right| \leq
10^{18}-10^{19}\,{\rm cm}^{2}$, the perihelion precession gives a
much stronger constraint $\left| Q\right| \leq 6\times
10^{7}-5\times 10^{8}\,{\rm cm}^{2}$. An improvement of one order
of magnitude in the observational data on Mercury's perihelion
shift could provide a very precise estimate of the bulk tidal
parameter.

Relative to the brane tension, the numerical value of $\lambda$
has been constrained by using big bang nucleosynthesis data, which
gives $\lambda \geq 1$ MeV$^{4}$ \cite{MaWa00}. A much stronger
constraint for the brane tension has been obtained by null results
of Newton's law at sub-millimeter scales. Bulk effects lead to the
modification of Newton's law on the brane.  The computation of the
Newton potential on the brane shows that the correction terms to
the Newton potential involve a logarithmic factor \cite{Ga00}.
When the distance between two point masses is very small with
respect to the AdS radius, the contribution of the Kaluza-Klein
spectrum becomes dominant as compared to the usual inverse square
law. This type of behavior of the Newtonian gravitational
potential may be used to prove the existence of an extra dimension
experimentally \cite{Ga00}. The null results of deviations from
Newton's law give the constraint $\lambda \geq 10^{8}$ GeV$^{4}$
\cite{MaWa00}.

Junction conditions relate the tidal parameter to the brane
tension and mass, and radius of a compact astrophysical object.
This can be used to calculate the numerical value of the brane
tension $\lambda$. However, despite the fact that these relations
have been derived for high constant density stars, for which
general relativistic effects are extremely strong (which is
definitely not the case for the Sun), their application to the
case of the solar system can give some approximate estimates of
$\lambda$. Stronger estimate, however, are implied from the
perihelion precession, giving $\lambda \geq 10^{5}\,{\rm
MeV}^{4}$. All estimates could be very much improved by the use of
observational data obtained for high density compact astrophysical
objects, like neutron stars.

On the other hand, there are several other vacuum solutions of the
spherically symmetric static gravitational field equations on the
brane \cite{branesolutions1}. Indeed, the effects due to the
projections of the Weyl tensor specify the deviations of brane
world models from general relativity. Since the generic form of
the Weyl tensor in the full five-dimensional theory is yet
unknown, the effects of known solutions must be studied on a case
by case basis. While one can in principle constrain the
projections, this only yields very mild constraints on the
five-dimensional Weyl tensor.

The study of the classical tests of the general relativity could
provide a very powerful method for constraining the allowed
parameter space of solutions, and to provide a deeper insight into
the physical nature and properties of the corresponding space-time
metrics. Therefore, this opens the possibility of testing brane
world models by using astronomical and astrophysical observations
at the solar system scale. Of course, this analysis must be
extended to the study of all vacuum solutions on the brane, which
requires developing general methods for the high precision study
of the classical tests in arbitrary spherically symmetric
space-times. In the present paper we have provided some basic
theoretical tools necessary for the in depth comparison of the
predictions of the brane world model and of the
observational/experimental results.

\ack We thank Roy Maartens for helpful discussions. The work of TH
was supported by the RGC grant No.~7027/06P of the government of
the Hong Kong SAR. FSNL was funded by Funda\c{c}\~{a}o para a
Ci\^{e}ncia e a Tecnologia (FCT)--Portugal through the grant
SFRH/BPD/26269/2006.

\section*{References}


\end{document}